\documentclass[preprint,12pt]{elsarticle}

\usepackage{colortbl}
\usepackage{amssymb}

\usepackage{amsmath}

\usepackage{amsthm}
\usepackage{xcolor}
\usepackage{soul}
\usepackage{graphicx}
\usepackage{hyperref}
\usepackage{natbib}

\journal{International Journal of Medical Informatics}

\begin{document}

\begin{frontmatter}

\title{Socio-Technical Risks of Clinical Speech-to-Text Systems: Transparency, Privacy, and Reliability Challenges in AI-Driven Documentation}

\author{Nelly Elsayed} 

\affiliation{organization={School of Information Technology, University of Cincinnati},
            state={Ohio},
            country={United States}}


\begin{abstract}

\textbf{Background:} 
AI-driven speech-to-text (STT) documentation systems are increasingly adopted in clinical settings to reduce documentation burden and improve workflow efficiency. However, adoption has outpaced the systematic evaluation of socio-technical risks related to transparency, reliability, patient autonomy, and organizational accountability.

\textbf{Objective:}
To develop a socio-technical framework for identifying and governing risks associated with the implementation of clinical speech-to-text systems.

\textbf{Methods:} 
This study synthesizes interdisciplinary evidence from technical automatic speech recognition research, clinical workflow and human factors studies, ethical guidance on consent and patient autonomy, and regulatory and organizational governance sources. Using a structured narrative synthesis approach, relevant literature was iteratively reviewed and thematically analyzed to identify recurring socio-technical risk mechanisms. The synthesis was used to develop a layered conceptual framework for evaluating and governing clinical speech-to-text systems.

\textbf{Results:} 
Findings show that clinical STT systems operate within tightly coupled socio-technical environments where model performance, audio capture conditions, clinician oversight, patient understanding, workflow design, and institutional governance are interdependent. Key risks include inconsistent disclosure and consent practices, performance disparities for accented speech and speech/voice disorders, accuracy degradation under real clinical acoustics, automation complacency and variable clinician review, and unclear accountability across vendors and healthcare organizations. These risk domains informed a six-layer socio-technical governance model spanning technical, human/workflow, ethical, organizational, regulatory, and sociocultural dimensions.

\textbf{Conclusion:} 
The study proposes a socio-technical governance framework and implementation roadmap to support the responsible deployment of clinical STT systems. The framework emphasizes transparency, patient autonomy, documentation integrity, and accountable governance to enable safe and equitable adoption of speech-based documentation technologies.

\end{abstract}




\end{frontmatter}



\section{Introduction}

Clinical documentation is a critical component of healthcare that shapes patient care progress, billing, clinical communication, and medico-legal accountability~\cite{suhail2025current,parekh2020documentation,sinsky_2016_workload,arndt_2017_ehrburden}. The rapid adoption of artificial intelligence (AI) systems has introduced a paradigm shift in clinical workflow and processes~\cite{maleki2024role}. AI-driven speech-to-text (STT) has been integrated into clinical documentation systems, changing how clinical notes are produced~\cite{goss2019clinician}. These systems aim to reduce documentation burden and improve efficiency by transcribing and summarizing clinical conversations using automatic speech recognition (ASR) and natural language processing (NLP), including large language model (LLM) summarization~\cite{suhail2025current,de2025improving,lee2023machine}.

The adoption of clinical STT systems has outpaced corresponding evidence and policy frameworks governing their transparency, accuracy, privacy protections, and equitable performance. Recent studies and emerging randomized evaluations report reductions in after-hours charting and improvements in clinician experience and documentation outcomes, supporting the promise of these tools in busy clinical settings~\cite{afshar2025ambient,lukac2025ambient,zuchowski2022speech,sasseville2025impact,sarraf2025impact}. However, adoption has outpaced systematic evaluation of socio-technical risks. In practice, patients may receive limited disclosure about recording, data handling, retention, and third-party access. Clinicians may over-trust outputs under time pressure. Organizations may rely on vendor benchmarks without independent validation. These concerns raise ethical and governance questions related to informed consent, autonomy, and patient rights, particularly when sensitive health conversations are captured and processed by external vendors or cloud-based AI systems~\cite{gerke2020ethical,longoni2019resistance,pool2022data}.

Clinical speech-to-text systems operate within what socio-technical systems theory describes as tightly coupled arrangements of technology, people, organizational structures, tasks, and regulatory environments~\cite{carayon2006work}. Socio-technical perspectives emphasize that technical performance cannot be separated from human practices, institutional governance, and broader social trust. Risks therefore emerge not only from model accuracy limitations, but also from workflow integration, consent practices, accountability structures, and cultural expectations surrounding AI use in healthcare.

To contextualize these risks, it is necessary to examine the current state of AI-driven speech documentation, the technical capabilities and limitations of modern ASR systems, and the emerging governance challenges surrounding their deployment.

\subsection{Growing Adoption of AI-Driven Speech Documentation}

Clinical documentation supports continuity of care, clinical decision-making, billing, legal protection, and interprofessional communication. With the expansion of electronic health records (EHRs), clinicians spend substantial time generating and editing notes, often exceeding time spent in direct patient care~\cite{sinsky_2016_workload,arndt_2017_ehrburden}. These administrative burdens contribute to burnout and reduced patient-facing time.

To address these challenges, healthcare organizations have increasingly adopted AI-driven STT systems and ambient documentation systems that transcribe clinician--patient conversations and generate structured visit notes using ASR and NLP/LLM summarization~\cite{goss2019clinician,min2023exploring}. Several studies report documentation time reductions and reduced after-hours charting, alongside improved clinician satisfaction~\cite{Duggan2025Clinician,zuchowski2022speech,sasseville2025impact,sarraf2025impact,guo2025evaluating}. More recently, randomized evaluations have begun to test these tools under pragmatic clinical conditions. Reported reductions in documentation time and after-hours charting have further accelerated institutional interest in adoption, while simultaneously highlighting the importance of evaluating safety, transparency, and governance implications alongside performance gains~\cite{afshar2025ambient,lukac2025ambient}.

\subsection{Technical Capabilities and Limitations of Clinical ASR}

Modern clinical speech systems rely on deep learning--based ASR models trained on large speech corpora, combined with language modeling and downstream NLP/LLM components to summarize dialogue into structured clinical notes~\cite{min2023exploring,alharbi2021automatic,bhardwaj2022automatic,padmanabhan2015machine}. While ASR accuracy has improved over time, reported benchmark performance does not necessarily reflect clinical environments, where acoustic variability, multi-speaker dialogue, and specialized terminology can degrade recognition and clinical utility~\cite{xiong2018microsoft,tae2019data,daneshjou2021lack}.

State-of-the-art ASR exhibits performance variability across speakers, environments, and linguistic characteristics. Clinical environments exacerbate these issues because conversations are spontaneous, involve multiple speakers, contain specialized terminology, and occur in acoustically challenging settings~\cite{kumar2024comprehensive}. ASR performance degrades substantially for atypical speech patterns, including dysarthric speech, stuttered speech, and neurodegenerative disease--related speech~\cite{rudzicz_2012_dysarthria,gambino_2021_stutter,martinez_2020_neuroasr}. Audio capture quality is also a significant determinant of reliability~\cite{fish2003using,gaur2016effects,gillespie2003strategies}. These constraints highlight a persistent gap: clinical ASR systems are often evaluated in controlled conditions rather than in noisy, dynamic environments where they are deployed in practice~\cite{daneshjou2021lack}.

\subsection{Real-World Reliability and Governance Challenges}

In addition to technical variability, clinical STT technologies introduce broader governance and privacy risks. Clinical conversations often contain highly sensitive information beyond the medical domain, including mental health concerns, family references, social determinants of health, and financial stressors. Audio and derived text may be accessible to third-party vendors, used for algorithmic improvement, or retained in cloud repositories outside direct organizational control, raising questions about HIPAA compliance, data minimization, cross-border transfers, retention, and security of identifiable speech data~\cite{vcartolovni2022ethical,kaplan2020revisiting,tayebi2024addressing,wiepert2022risk}.

Despite the large-scale deployment of STT tools, there is currently no unified socio-technical framework guiding their implementation, auditing, and communication to patients. Healthcare organizations lack standardized requirements for performance monitoring, bias evaluation, transparency practices, clinician oversight, and patient rights, including the right to opt out or use alternative documentation methods. As a result, AI-driven documentation systems operate in an environment of unclear accountability, variable clinical governance, and significant ethical uncertainty.

Collectively, these gaps in transparency, reliability evaluation, and governance demonstrate that clinical STT implementation raises interconnected socio-technical risks that cannot be addressed through technical validation alone. A structured analysis is therefore needed to clarify how these risks arise, how they interact across domains, and what governance mechanisms are required to mitigate them. To address these issues, this paper investigates the following research questions:

\begin{itemize}
    \item \textbf{\textit{RQ1:}} What transparency and consent practices are used when clinical STT systems document patient–provider interactions, and how do these practices shape patient understanding and autonomy?
    \item \textit{\textbf{RQ2:}} How do real-world clinical acoustics and speech diversity affect STT reliability, and what documentation-quality and safety implications follow?
    \item \textbf{\textit{RQ3:}} What privacy, data governance, and accountability issues arise from recording, processing, and storing clinical conversations, and what socio-technical governance mechanisms can mitigate them?
\end{itemize}

By addressing these questions, this paper contributes a structured socio-technical analysis of the risks embedded in AI-driven clinical documentation and proposes governance mechanisms to support safe, transparent, and accountable implementation.
\section{Methods: Structured Narrative Evidence Synthesis}

This study employed a structured narrative synthesis to integrate evidence across technical, clinical, ethical, and regulatory domains relevant to AI-driven clinical speech documentation.

This work was not designed as a systematic review aimed at exhaustive identification of all published studies. Rather, it used a structured and transparent literature identification process to support conceptual framework development and thematic synthesis of socio-technical risk domains.

\subsection{Scope and Design}

The review was conceptual and integrative rather than empirical, aiming to identify recurrent risk patterns and governance gaps across disciplines and to synthesize these into a coherent socio-technical framework.

\subsection{Source Identification}

Literature was identified from PubMed/MEDLINE, Scopus, IEEE Xplore, the ACM Digital Library, and ScienceDirect, supplemented by policy and regulatory sources (e.g., HIPAA guidance, GDPR documentation, and the NIST AI Risk Management Framework).

The search period was defined from January 1, 2015 to December 31, 2025. The year 2015 was selected as the starting point because it corresponds to the rapid maturation and clinical translation of deep learning–based ASR and the subsequent integration of neural network architectures into healthcare documentation workflows~\cite{padmanabhan2015machine}. The final search update was conducted in January 2026.

Searches used combinations of terms including \emph{clinical speech-to-text}, \emph{ambient AI scribe}, \emph{ASR healthcare}, \emph{informed consent}, \emph{privacy}, and \emph{AI governance}. In addition to database searches, backward and forward citation tracking was performed to identify relevant foundational and emerging studies.

\subsection{Inclusion and Exclusion Criteria}

Included sources addressed one or more of the following domains: (a) clinical deployment or evaluation of STT systems, (b) ASR performance under real-world or diverse speech conditions, (c) workflow and human factors implications, (d) ethical and consent considerations, or (e) regulatory and organizational governance.

Sources focused exclusively on non-clinical dictation systems, consumer voice assistants, or purely technical ASR optimization without relevance to clinical documentation workflows were excluded. Perspective/commentary pieces were generally excluded unless they provided were also excluded unless they provided regulatory or policy guidance relevant to governance.

\subsection{Thematic Synthesis and Framework Development}

Relevant findings were extracted and iteratively reviewed using manual thematic coding. Themes were identified based on recurring descriptions of risk mechanisms, implementation challenges, oversight gaps, and governance considerations across the literature. These themes were then grouped into higher-level socio-technical domains reflecting technical infrastructure, human interaction and workflow, ethical and transparency practices, organizational governance, legal and regulatory compliance, and sociocultural trust.

The socio-technical layers presented in this study were derived from this thematic clustering process rather than imposed a priori. Governance recommendations were subsequently mapped to each layer by linking identified risk mechanisms to corresponding control strategies reported in the literature or proposed in regulatory and organizational guidance.

Across the reviewed literature, five dominant thematic clusters emerged. First, technical performance variability, including acoustic sensitivity, speech diversity disparities, and domain adaptation limitations. Second, workflow integration challenges, particularly clinician review burden, automation complacency, and error propagation into EHR systems. Third, transparency and informed consent variability, with inconsistent disclosure practices and limited patient awareness of data processing pathways. Fourth, data governance and privacy concerns, including retention policies, third-party vendor access, model retraining use, and re-identification risks. Fifth, accountability ambiguity, reflecting unclear responsibility boundaries among vendors, clinicians, and healthcare organizations. These recurring themes informed the construction of the six-layer socio-technical framework presented in this study.


\section{Problem Analysis}\label{sec:ProblemAnalysis}

\subsection{Transparency and Informed Consent Challenges (RQ1)}
In many clinical settings, disclosure is limited to general statements that recording ``assists documentation,'' without meaningful detail about processing location, retention, third-party access, or model-training use~\cite{ng2025evaluating,mcgraw2021privacy}. This weakens informed consent and patient autonomy, particularly when patients cannot easily opt out without affecting care experience~\cite{entwistle2010supporting,xafis2019ethics}. Recent evidence shows variability in consent approaches and stakeholder expectations regarding ambient documentation, reinforcing the need for standardized, patient-centered disclosure and opt-out pathways~\cite{lawrence2025consent,topaz2025beyond,rockwern2021health}.

\subsection{Reliability Risks in Real-World Clinical Settings (RQ2)}
Clinical environments introduce acoustic conditions that degrade ASR accuracy, including HVAC systems, hallway activity, alarms, medical equipment, reverberation, and echo. Empirical research shows that increases in distance or microphone orientation can materially reduce ASR accuracy, and relatively modest background noise shifts can distort transcription quality~\cite{pearce2000aurora,johnson2014systematic}. These conditions affect typical speech and disproportionately impact patients with speech diversity, including accented speech, multilingual speech, and speech disorders~\cite{jefferson2019usability,errattahi2018automatic,young2010difficulties,sahu2017challenges,rudzicz_2012_dysarthria,gambino_2021_stutter,martinez_2020_neuroasr}. Because speech and voice conditions are common, failure to validate systems for these profiles introduces equity concerns and systematic documentation quality differences~\cite{nidcd_stats_voice}.

\subsection{Automation Complacency and Variable Clinician Oversight (RQ2)}
As a downstream consequence of reliability limitations discussed in RQ2, clinician trust in automated outputs can lead to reduced review and automation complacency.
When automated notes appear ``good enough,'' time pressure can reduce review depth and increase the probability that transcription errors persist in the EHR~\cite{lyell2017automation}. Documentation errors may then propagate across encounters via copy-forward behaviors and downstream use for clinical decision-making, billing, and medico-legal documentation~\cite{zhou2018analysis,goss2016incidence}.

\subsection{Privacy, Data Governance, and Accountability Gaps (RQ3)}
Clinical conversations include highly sensitive information beyond strictly medical content (e.g., mental health, family, finances, social determinants). Audio and derived text may be stored or processed by vendors, potentially implicating retention, access controls, secondary use, and cross-border data flows~\cite{vcartolovni2022ethical,kaplan2020revisiting}. Voice biometrics and background audio increase re-identification risks~\cite{tayebi2024addressing,wiepert2022risk,koffi2023voice}. Accountability for errors and incidents is often unclear: clinicians bear legal responsibility for note accuracy, while vendors control system design, model updates, and data handling, complicating auditing, remediation, and patient recourse~\cite{mcgraw2021privacy,topaz2025beyond}.

\section{Socio-Technical Conceptual Framework}\label{sec:framework}

The challenges identified above indicate that clinical STT systems cannot be evaluated or governed solely as technical tools. A socio-technical perspective supports systematic examination of how technical performance, workflow, ethics, governance, regulation, and trust interact to shape documentation outcomes.

Unlike linear risk models, the framework emphasizes \textit{bidirectional coupling} between layers, where governance, ethical practices, and sociocultural trust both shape and are shaped by technical performance and clinical workflow. Failures or interventions at any layer may therefore propagate upward or downward through feedback loops, amplifying or mitigating risks to documentation accuracy, patient autonomy, and institutional trust.

\begin{figure}[t]
    \centering
    \includegraphics[width=5cm, height= 7 cm]{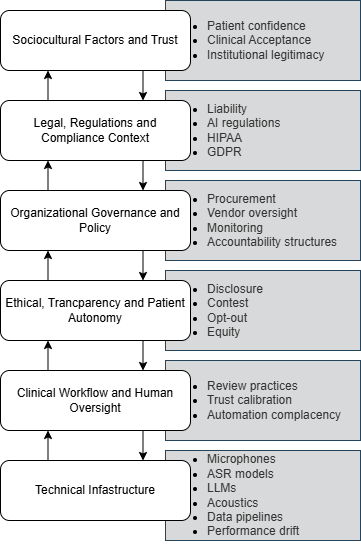}
    \caption{Socio-technical framework for clinical speech-to-text documentation. The framework depicts layered but \textit{bidirectionally coupled} technical, human, ethical, organizational, regulatory, and sociocultural factors. Risks and failures may originate at any layer and propagate across others through feedback loops, shaping documentation quality, patient autonomy, and institutional trust.}
    \label{fig:socio_technical_framework}
\end{figure}

Figure~\ref{fig:socio_technical_framework} presents six coupled layers for responsible clinical speech documentation: technical infrastructure, human interaction and clinical workflow, ethical/transparency and patient autonomy, organizational governance and policy, legal/regulatory compliance, and sociocultural trust.
The vertical arrangement of layers in Figure~\ref{fig:socio_technical_framework} does not imply a hierarchy of importance or a linear progression of risk. Rather, the positioning reflects increasing levels of abstraction—from immediate technical infrastructure and clinical workflow processes to broader governance, regulatory, and sociocultural contexts. Each layer is analytically distinct but functionally interdependent. Technical performance influences trust and governance, while regulatory requirements, organizational policy, and cultural expectations simultaneously constrain and shape technical system design and clinical use.

\subsection{Technical Infrastructure Layer}
This layer includes ASR models, microphones and capture devices, acoustic conditions, including diarization and downstream NLP/LLM summarization. Variability in audio capture and speaker diversity can produce clinically meaningful errors that propagate into documentation~\cite{daneshjou2021lack,rudzicz_2012_dysarthria}.

\subsection{Human Interaction and Clinical Workflow Layer}
This layer governs how clinicians and patients interact with STT tools during care. Review practices, time pressure, interface design, and trust calibration determine whether errors are detected and corrected~\cite{lyell2017automation,zhou2018analysis}.

\subsection{Ethical, Transparency, and Patient Autonomy Layer}
Ethical deployment requires meaningful disclosure, explicit consent, and feasible opt-out pathways. Equity considerations are central because differential ASR performance can create systematically lower documentation quality for some patient groups~\cite{mcgraw2021privacy,lawrence2025consent}.

\subsection{Organizational Governance and Policy Layer}
This layer includes procurement, vendor oversight, internal policies, performance monitoring, escalation pathways, and incident response. Weak governance increases the likelihood that risks remain unmeasured and unmanaged~\cite{topaz2025beyond,kaplan2020revisiting}.

\subsection{Legal, Regulatory, and Compliance Layer}
Regulatory obligations (e.g., HIPAA; GDPR, where applicable) shape requirements for access controls, data minimization, retention/deletion, contracting, and breach response~\cite{vcartolovni2022ethical,tayebi2024addressing}. Ambiguities for real-time audio capture and AI summarization require organizations to translate high-level rules into operational controls.

\subsection{Sociocultural and Trust Layer}
Trust is an emergent outcome shaped by transparency, reliability, governance, and cultural expectations~\cite{longoni2019resistance,gerke2020ethical}. Loss of trust can reduce patient candor and clinician engagement, indirectly harming both communication quality and system performance.

\section{Governance Strategies and Recommendations}\label{sec:governance}

To strengthen coherence between the framework and recommendations, we map each layer to key risks and representative governance controls.

A structured mapping of socio-technical layers to key risks and representative governance controls is presented in Table~\ref{tab:layer_mapping} to support practical implementation and cross-layer alignment.

\begin{table}[t]
\centering
\caption{Mapping socio-technical layers to representative risks and governance controls for clinical STT systems.}
\label{tab:layer_mapping}
\begin{tabular}{p{2.6cm} p{5.2cm} p{5.2cm}}
\hline
\textbf{Layer} & \textbf{Key Risks} & \textbf{Governance Controls} \\
\hline

Technical Infrastructure 
& Acoustic variability \newline
  Speech diversity disparities \newline
  Model drift
& Local validation \newline
  Equity testing \newline
  Continuous performance monitoring \\

Human Interaction / Workflow 
& Automation complacency \newline
  Incomplete review \newline
  Error propagation
& Human-in-the-loop standards \newline
  Clinician training \newline
  Structured error reporting \\

Ethical / Transparency 
& Limited disclosure \newline
  Constrained opt-out \newline
  Autonomy concerns
& Standardized patient disclosure \newline
  Explicit consent pathways \newline
  Alternative documentation options \\

Organizational Governance 
& Vendor opacity \newline
  Unclear accountability \newline
  Oversight gaps
& Contractual data controls \newline
  Accountability matrices \newline
  Multidisciplinary oversight committees \\

Legal / Regulatory 
& Retention ambiguity \newline
  Access control gaps \newline
  Compliance risks
& Data minimization \newline
  Defined retention schedules \newline
  Logging and incident response protocols \\

Sociocultural Trust 
& Reduced patient candor \newline
  Legitimacy concerns \newline
  Trust erosion
& Transparent communication \newline
  Trust monitoring \newline
  Governance-performance feedback loops \\

\hline
\end{tabular}
\end{table}

\color{black}{}

\subsection{Technical Reliability and Performance Controls}
\begin{itemize}
    \item \textit{Local validation:} Require pre-deployment testing under local clinical acoustics (room type, device placement, multi-speaker conditions) and clinical vocabulary.
    \item \textit{Equity testing:} Evaluate performance across accents, multilingual speech, and speech/voice disorders using representative scenarios.
    \item \textit{Continuous monitoring:} Implement monitoring for drift, error patterns, and failure modes (e.g., negation, medication entities), with thresholds that trigger retraining, configuration changes, or workflow safeguards.
\end{itemize}

\subsection{Workflow and Human Oversight Controls}
\begin{itemize}
    \item \textit{Human-in-the-loop standards:} Define minimum review requirements (e.g., review of medication list, problems, plan, and key negatives) before signing notes.
    \item \textit{Training:} Train clinicians on common ASR failure modes and automation complacency risks; provide quick correction workflows.
    \item \textit{Error reporting:} Provide low-friction mechanisms to flag errors and route issues to governance teams and vendors.
\end{itemize}

\subsection{Ethical, Transparency, and Autonomy Controls}
\begin{itemize}
    \item \textit{Standardized disclosure:} Provide patient-facing explanations describing what is recorded, where it is processed, retention, third-party access, and any model-improvement uses.
    \item \textit{Consent and opt-out:} Use explicit consent processes appropriate to the context and ensure meaningful opt-out pathways without penalizing care.
    \item \textit{Equity safeguards:} Offer alternatives (e.g., manual documentation, human scribe) when speech differences or environmental conditions reduce reliability.
\end{itemize}

\subsection{Organizational Governance and Accountability}
\begin{itemize}
    \item \textit{Vendor oversight:} Require contractual clarity on data handling, retention/deletion, training use, audit rights, and incident response.
    \item \textit{Accountability matrix:} Define responsibilities for performance monitoring, corrections, incident management, and patient communication.
    \item \textit{Safety governance:} Establish a multidisciplinary oversight committee (clinical leadership, IT, compliance, privacy, and patient advocacy).
\end{itemize}

\subsection{Legal and Regulatory Compliance Controls}
\begin{itemize}
    \item \textit{Data minimization and retention:} Limit retention of raw audio where feasible, define retention schedules, and deletion verification.
    \item \textit{Access controls:} Apply least-privilege access, logging, and periodic access review for audio and transcripts.
    \item \textit{Breach preparedness:} Maintain incident response plans tailored to audio capture and vendor-managed systems.
\end{itemize}

\subsection{Implementation Roadmap (Phased Deployment)}
A responsible deployment roadmap includes: (1) readiness assessment (governance, privacy, workflows), (2) vendor evaluation and contracting, (3) pilot deployment with local validation and equity testing, (4) clinician training and patient-facing disclosure rollout, (5) phased scaling with monitoring dashboards, and (6) continuous improvement through audits, feedback loops, and incident reviews.

\section{Discussion}

This synthesis reinforces that clinical STT systems are safety- and trust-sensitive technologies embedded in complex socio-technical environments. Governance must extend beyond aggregate model accuracy to address transparency, consent, and patient agency, workflow oversight, equity impacts, and vendor accountability. The proposed framework provides a structure for analyzing risk propagation and aligning controls with the specific mechanisms that generate documentation risk.

\section{Conclusion}

AI-driven STT systems can reduce documentation burden, but their responsible use requires socio-technical governance that integrates technical validation with workflow safeguards, patient-centered transparency, robust privacy controls, and clear accountability across organizations and vendors. The proposed framework and roadmap provide actionable guidance for safe and equitable clinical adoption.

This study has several limitations. First, the analysis is conceptual and does not report empirical findings from a single deployment or controlled evaluation. Second, although the structured narrative synthesis was designed to identify recurrent risk mechanisms, it does not provide exhaustive coverage of all published studies. Finally, future research should empirically evaluate governance interventions in real-world clinical settings, assess equity impacts across diverse patient populations, and examine how socio-technical feedback loops evolve over time as clinical speech-to-text systems mature.



\section{Submission Declaration}
\begin{itemize}
    \item The work described has not been published previously except in the form of a preprint, an abstract, a published lecture, academic thesis or registered report.
    \item The article's publication is approved by all authors and tacitly or explicitly by the responsible authorities where the work was carried out.
    \item If accepted, the article will not be published elsewhere in the same form, in English or in any other language, including electronically, without the written consent of the copyright-holder.
\end{itemize}

\section{Authorship}
All authors (sole author) have made substantial contributions including conception and design of the study, drafting the article and perform fully revision of the content.

\section{Declaration of Interests}
The author declare that they have no known competing financial interests or personal relationships that could have appeared to influence the work reported in this paper.

\section{Funding}
This research did not receive any specific grant from funding agencies in the public, commercial, or not-for-profit sectors.

\section{Declaration of Generative AI Use}
During the preparation of this work the author(s) used ChatGPT, Gemmini, Google Scholar Lab, and Grammarly in order to assist in draft refining, and organizing sections of the manuscript, spell-check and grammar-check, and Latex syntax revision. The author reviewed, validated, and edited all content to ensure accuracy, integrity, and alignment with scholarly standards and take(s) full responsibility for the content of the published article.


\section{Data Statement}
No datasets were generated or analyzed for this study. This work is a conceptual and theoretical analysis based solely on publicly available literature and does not involve clinical data, human subjects, or proprietary datasets. Therefore, no data are associated with this manuscript.

 \bibliographystyle{elsarticle-num} 
\bibliography{references_speech}

@article{sinsky_2016_workload,
  title={Allocation of physician time in ambulatory practice: a time and motion study in 4 specialties},
  author={Sinsky, Christine and Colligan, Lacey and Li, Lin and Prgomet, Mirela and Reynolds, Sandra and Goeders, Leigh and Westbrook, Johanna},
  journal={Annals of Internal Medicine},
  volume={165},
  number={11},
  pages={753--760},
  year={2016},
  publisher={American College of Physicians}
}

@article{padmanabhan2015machine,
  title={Machine learning in automatic speech recognition: A survey},
  author={Padmanabhan, Jayashree and Johnson Premkumar, Melvin Jose},
  journal={IETE Technical Review},
  volume={32},
  number={4},
  pages={240--251},
  year={2015},
  publisher={Taylor \& Francis}
}

@article{carayon2006work,
  title={Work system design for patient safety: the SEIPS model},
  author={Carayon, PASH and Hundt, A Schoofs and Karsh, Ben-Tzion and Gurses, Ayse P and Alvarado, Carla J and Smith, Michael and Brennan, P Flatley},
  journal={BMJ Quality \& Safety},
  volume={15},
  number={suppl 1},
  pages={i50--i58},
  year={2006},
  publisher={BMJ Publishing Group Ltd}
}

@article{arndt_2017_ehrburden,
  title={Tethered to the {EHR}: Primary care physician workload assessment using EHR event log data and time-motion observations},
  author={Arndt, Brian G. and Beasley, Jean W. and Watkinson, Mark D. and Temte, Jon L. and Tuan, Wen-Juo and Sinsky, Christine A. and Gilchrist, V. James},
  journal={Annals of Family Medicine},
  volume={15},
  number={5},
  pages={419--426},
  year={2017}
}

@article{afshar2025ambient,
  title={A pragmatic randomized controlled trial of ambient artificial intelligence to improve health practitioner well-being},
  author={Afshar, Majid and Ryan Baumann, Mary and Resnik, Felice and Hintzke, Josie and Gravel Sullivan, Anne and Wills, Graham and Lemmon, Kayla and Dambach, Jason and Mrotek, Leigh Ann and Quinn, Mariah and others},
  journal={NEJM AI},
  volume={2},
  number={12},
  pages={AIoa2500945},
  year={2025},
  publisher={Massachusetts Medical Society}
}

@article{lukac2025ambient,
  title={Ambient AI scribes in clinical practice: A randomized trial},
  author={Lukac, Paul J and Turner, William and Vangala, Sitaram and Chin, Aaron T and Khalili, Joshua and Shih, Ya-Chen Tina and Sarkisian, Catherine and Cheng, Eric M and Mafi, John N},
  journal={NEJM AI},
  volume={2},
  number={12},
  pages={AIoa2501000},
  year={2025},
  publisher={Massachusetts Medical Society}
}

@article{lawrence2025consent,
  title={Informed consent for ambient documentation using generative AI in ambulatory care},
  author={Lawrence, Katharine and Kuram, Vasudev S and Levine, Defne L and Sharif, Sarah and Polet, Conner and Malhotra, Kiran and Owens, Kellie},
  journal={JAMA network open},
  volume={8},
  number={7},
  pages={e2522400--e2522400},
  year={2025},
  publisher={American Medical Association}
}

@article{Duggan2025Clinician,
  author  = {Duggan, Matthew J and Gervase, Julietta and Schoenbaum, Anna and Hanson, William and Howell, John T and Sheinberg, Michael and Johnson, Kevin B},
  title   = {Clinician Experiences With Ambient Scribe Technology to Assist With Documentation Burden and Efficiency},
  journal = {JAMA Netw Open},
  year    = {2025},
  volume  = {8},
  number  = {2},
  pages   = {e2460637},
  doi     = {10.1001/jamanetworkopen.2024.60637},
  pmcid   = {PMC11840636}
}

@inproceedings{sasseville2025impact,
  title={The Impact of AI Scribes on Streamlining Clinical Documentation: A Systematic Review},
  author={Sasseville, Maxime and Yousefi, Farzaneh and Ouellet, Steven and Naye, Florian and Stefan, Th{\'e}o and Carnovale, Val{\'e}rie and Bergeron, Fr{\'e}d{\'e}ric and Ling, Linda and Gheorghiu, Bobby and Hagens, Simon and others},
  booktitle={Healthcare},
  volume={13},
  number={12},
  pages={1447},
  year={2025}
}

@article{guo2025evaluating,
  title={Evaluating ambient artificial intelligence documentation: effects on work efficiency, documentation burden, and patient-centered care},
  author={Guo, Yawen and Wang, Jiayuan and Hu, Di and Tam, Steven and Gilman, Charles and Chow, Emilie and Perret, Danielle and Pandita, Deepti and Zheng, Kai},
  journal={Journal of the American Medical Informatics Association},
  pages={ocaf180},
  year={2025},
  publisher={Oxford University Press}
}

@article{zuchowski2022speech,
  title={Speech recognition for medical documentation: an analysis of time, cost efficiency and acceptance in a clinical setting},
  author={Zuchowski, Matthias and G{\"o}ller, Aydan},
  journal={British Journal of Healthcare Management},
  volume={28},
  pages={30--36},
  year={2022},
  publisher={MA Healthcare London}
}

@article{sarraf2025impact,
  title={Impact of artificial intelligence on electronic health record-related burnouts among healthcare professionals: systematic review},
  author={Sarraf, Berna and Ghasempour, Ali},
  journal={Frontiers in Public Health},
  volume={13},
  pages={1628831},
  year={2025},
  publisher={Frontiers}
}

@inproceedings{xiong2018microsoft,
  title={The Microsoft 2017 conversational speech recognition system},
  author={Xiong, Wayne and Wu, Lingfeng and Alleva, Fil and Droppo, Jasha and Huang, Xuedong and Stolcke, Andreas},
  booktitle={2018 IEEE international conference on acoustics, speech and signal processing (ICASSP)},
  pages={5934--5938},
  year={2018},
  organization={IEEE}
}

@article{parekh2020documentation,
  title={Documentation in healthcare: Standards and guidelines},
  author={Parekh, Utsav},
  journal={Legal Issues in Medical Practice},
  volume={145},
  year={2020},
  publisher={Jaypee Brothers Medical Publishers}
}

@incollection{gerke2020ethical,
  title={Ethical and legal challenges of artificial intelligence-driven healthcare},
  author={Gerke, Sara and Minssen, Timo and Cohen, Glenn},
  booktitle={Artificial intelligence in healthcare},
  pages={295--336},
  year={2020},
  publisher={Elsevier}
}

@article{longoni2019resistance,
  title={Resistance to medical artificial intelligence},
  author={Longoni, Chiara and Bonezzi, Andrea and Morewedge, Carey K},
  journal={Journal of consumer research},
  volume={46},
  number={4},
  pages={629--650},
  year={2019},
  publisher={Oxford University Press}
}

@article{pool2022data,
  title={Data privacy concerns and use of telehealth in the aged care context: an integrative review and research agenda},
  author={Pool, Javad and Akhlaghpour, Saeed and Fatehi, Farhad and Gray, Leonard C},
  journal={International Journal of Medical Informatics},
  volume={160},
  pages={104707},
  year={2022},
  publisher={Elsevier}
}

@misc{nidcd_stats_voice,
  author = {{National Institute on Deafness and Other Communication Disorders}},
  title = {Quick Statistics About Voice, Speech, Language},
  year = {2024},
  url = {https://www.nidcd.nih.gov/health/statistics/quick-statistics-voice-speech-language},
  note = {Accessed: 2025-11-28}
}

@article{vcartolovni2022ethical,
  title={Ethical, legal, and social considerations of AI-based medical decision-support tools: a scoping review},
  author={{\v{C}}artolovni, Anto and Tomi{\v{c}}i{\'c}, Ana and Mosler, Elvira Lazi{\'c}},
  journal={International Journal of Medical Informatics},
  volume={161},
  pages={104738},
  year={2022},
  publisher={Elsevier}
}

@article{kaplan2020revisiting,
  title={Revisiting health information technology ethical, legal, and social issues and evaluation: telehealth/telemedicine and COVID-19},
  author={Kaplan, Bonnie},
  journal={International journal of medical informatics},
  volume={143},
  pages={104239},
  year={2020},
  publisher={Elsevier}
}

@article{alharbi2021automatic,
  title={Automatic speech recognition: Systematic literature review},
  author={Alharbi, Sadeen and Alrazgan, Muna and Alrashed, Alanoud and Alnomasi, Turkiayh and Almojel, Raghad and Alharbi, Rimah and Alharbi, Saja and Alturki, Sahar and Alshehri, Fatimah and Almojil, Maha},
  journal={Ieee Access},
  volume={9},
  pages={131858--131876},
  year={2021},
  publisher={IEEE}
}

@misc{fish2003using,
  title={Using audio quality to predict word error rate in an automatic speech recognition system},
  author={Fish, Randall and Hu, Qian and Boykin, Stanley},
  year={2003}
}

@article{kumar2024comprehensive,
  title={A comprehensive analysis of speech recognition systems in healthcare: current research challenges and future prospects},
  author={Kumar, Yogesh},
  journal={SN Computer Science},
  volume={5},
  number={1},
  pages={137},
  year={2024},
  publisher={Springer}
}

@inproceedings{gaur2016effects,
  title={The effects of automatic speech recognition quality on human transcription latency},
  author={Gaur, Yashesh and Lasecki, Walter S and Metze, Florian and Bigham, Jeffrey P},
  booktitle={Proceedings of the 13th International Web for All Conference},
  pages={1--8},
  year={2016}
}

@inproceedings{gillespie2003strategies,
  title={Strategies for improving audible quality and speech recognition accuracy of reverberant speech},
  author={Gillespie, Bradford W and Atlas, AE},
  booktitle={2003 IEEE International Conference on Acoustics, Speech, and Signal Processing, 2003. Proceedings.(ICASSP'03).},
  volume={1},
  pages={I--I},
  year={2003},
  organization={IEEE}
}

@inproceedings{min2023exploring,
  title={Exploring the integration of large language models into automatic speech recognition systems: An empirical study},
  author={Min, Zeping and Wang, Jinbo},
  booktitle={International Conference on Neural Information Processing},
  pages={69--84},
  year={2023},
  organization={Springer}
}

@article{bhardwaj2022automatic,
  title={Automatic speech recognition (asr) systems for children: A systematic literature review},
  author={Bhardwaj, Vivek and Ben Othman, Mohamed Tahar and Kukreja, Vinay and Belkhier, Youcef and Bajaj, Mohit and Goud, B Srikanth and Rehman, Ateeq Ur and Shafiq, Muhammad and Hamam, Habib},
  journal={Applied Sciences},
  volume={12},
  number={9},
  pages={4419},
  year={2022},
  publisher={MDPI}
}

@article{tayebi2024addressing,
  title={Addressing challenges in speaker anonymization to maintain utility while ensuring privacy of pathological speech},
  author={Tayebi Arasteh, Soroosh and Arias-Vergara, Tom{\'a}s and P{\'e}rez-Toro, Paula Andrea and Weise, Tobias and Packh{\"a}user, Kai and Schuster, Maria and Noeth, Elmar and Maier, Andreas and Yang, Seung Hee},
  journal={Communications Medicine},
  volume={4},
  number={1},
  pages={182},
  year={2024},
  publisher={Nature Publishing Group UK London}
}

@article{wiepert2022risk,
  title={Risk of re-identification for shared clinical speech recordings},
  author={Wiepert, Daniela A and Malin, Bradley A and Duffy, Joseph R and Utianski, Rene L and Stricker, John L and Jones, David T and Botha, Hugo},
  journal={arXiv preprint arXiv:2210.09975},
  year={2022}
}

@article{johnson2014systematic,
  title={A systematic review of speech recognition technology in health care},
  author={Johnson, Maree and Lapkin, Samuel and Long, Vanessa and Sanchez, Paula and Suominen, Hanna and Basilakis, Jim and Dawson, Linda},
  journal={BMC medical informatics and decision making},
  volume={14},
  number={1},
  pages={94},
  year={2014},
  publisher={Springer}
}

@misc{jefferson2019usability,
  title={Usability of Automatic Speech Recognition Systems for Individuals with Speech Disorders: Past, Present, Future, and A Proposed Model},
  author={Madeline Jefferson},
  year={2019},
  url={https://api.semanticscholar.org/CorpusID:196178405}
}

@article{sahu2017challenges,
  title={Challenges and issues in adopting speech recognition},
  author={Sahu, Priyanka and Dua, Mohit and Kumar, Ankit},
  journal={Speech and Language Processing for Human-Machine Communications: Proceedings of CSI 2015},
  pages={209--215},
  year={2017},
  publisher={Springer}
}

@article{young2010difficulties,
  title={Difficulties in automatic speech recognition of dysarthric speakers and implications for speech-based applications used by the elderly: A literature review},
  author={Young, Victoria and Mihailidis, Alex},
  journal={Assistive Technology},
  volume={22},
  number={2},
  pages={99--112},
  year={2010},
  publisher={Taylor \& Francis}
}

@article{errattahi2018automatic,
  title={Automatic speech recognition errors detection and correction: A review},
  author={Errattahi, Rahhal and El Hannani, Asmaa and Ouahmane, Hassan},
  journal={Procedia Computer Science},
  volume={128},
  pages={32--37},
  year={2018},
  publisher={Elsevier}
}

@article{lee2023machine,
  title={Machine learning-based speech recognition system for nursing documentation--A pilot study},
  author={Lee, Tso-Ying and Li, Chin-Ching and Chou, Kuei-Ru and Chung, Min-Huey and Hsiao, Shu-Tai and Guo, Shu-Liu and Hung, Lung-Yun and Wu, Hao-Ting},
  journal={International Journal of Medical Informatics},
  volume={178},
  pages={105213},
  year={2023},
  publisher={Elsevier}
}

@article{maleki2024role,
  title={The role of AI in hospitals and clinics: transforming healthcare in the 21st century},
  author={Maleki Varnosfaderani, Shiva and Forouzanfar, Mohamad},
  journal={Bioengineering},
  volume={11},
  number={4},
  pages={337},
  year={2024},
  publisher={MDPI}
}

@article{de2025improving,
  title={Improving documentation quality and patient interaction with AI: a tool for transforming medical records—an experience report},
  author={de Paula, Pedro Angelo Basei and Severino, Jo{\~a}o Victor Bruneti and Berger, Matheus Nespolo and Veiga, Maria Han and Ribeiro, Karen Dyminski Parente and Loures, Fillipe Silveira and Todeschini, Solano Amadori and Roeder, Eduardo Augusto and Marques, Gustavo Lenci},
  journal={Journal of Medical Artificial Intelligence},
  volume={8},
  year={2025},
  publisher={AME Publishing Company}
}

@article{suhail2025current,
  title={Current evidence and future directions of metrics used to evaluate ambient clinical documentation: A scoping review},
  author={Suhail, Dernas and Wong, Zhen Yu and Ubhi, Jeevan and Kungwengwe, Garikai and Faderani, Ryan and Mosahebi, Afshin},
  journal={International Journal of Medical Informatics},
  pages={106113},
  year={2025},
  publisher={Elsevier}
}

@article{goss2019clinician,
  title={A clinician survey of using speech recognition for clinical documentation in the electronic health record},
  author={Goss, Foster R and Blackley, Suzanne V and Ortega, Carlos A and Kowalski, Leigh T and Landman, Adam B and Lin, Chen-Tan and Meteer, Marie and Bakes, Samantha and Gradwohl, Stephen C and Bates, David W and others},
  journal={International journal of medical informatics},
  volume={130},
  pages={103938},
  year={2019},
  publisher={Elsevier}
}

@article{rudzicz_2012_dysarthria,
  title={Adaptive, personalized speech recognition for dysarthric speakers},
  author={Rudzicz, Frank},
  journal={IEEE Transactions on Audio, Speech, and Language Processing},
  volume={20},
  number={2},
  pages={544--555},
  year={2012}
}

@article{gambino_2021_stutter,
  title={Improving speech recognition for stuttered speech: A deep learning approach},
  author={Gambino, Michael and Patel, Manan},
  journal={Interspeech},
  pages={128--132},
  year={2021}
}

@article{martinez_2020_neuroasr,
  title={Automatic speech recognition for neurodegenerative disease: Challenges and opportunities},
  author={Martinez, Diego and Rosen, Lisa},
  journal={Journal of Speech, Language, and Hearing Research},
  volume={63},
  number={7},
  pages={2059--2073},
  year={2020}
}

@article{koffi2023voice,
  title={Voice biometrics fusion for enhanced security and speaker recognition: A comprehensive review},
  author={Koffi, Ettien},
  journal={Linguistic Portfolios},
  volume={12},
  number={1},
  pages={6},
  year={2023}
}

@article{topaz2025beyond,
  title={Beyond human ears: navigating the uncharted risks of AI scribes in clinical practice},
  author={Topaz, Maxim and Peltonen, Laura Maria and Zhang, Zhihong},
  journal={npj Digital Medicine},
  volume={8},
  number={1},
  pages={569},
  year={2025},
  publisher={Nature Publishing Group UK London}
}

@article{mcgraw2021privacy,
  title={Privacy protections to encourage use of health-relevant digital data in a learning health system},
  author={McGraw, Deven and Mandl, Kenneth D},
  journal={NPJ digital medicine},
  volume={4},
  number={1},
  pages={2},
  year={2021},
  publisher={Nature Publishing Group UK London}
}

@inproceedings{pearce2000aurora,
  title={The aurora experimental framework for the performance evaluation of speech recognition systems under noisy conditions.},
  author={Pearce, David and Hirsch, Hans-G{\"u}nter and others},
  booktitle={Interspeech},
  pages={29--32},
  year={2000}
}

@article{lyell2017automation,
  title={Automation bias and verification complexity: a systematic review},
  author={Lyell, David and Coiera, Enrico},
  journal={Journal of the American Medical Informatics Association},
  volume={24},
  number={2},
  pages={423--431},
  year={2017},
  publisher={Oxford University Press}
}

@article{zhou2018analysis,
  title={Analysis of errors in dictated clinical documents assisted by speech recognition software and professional transcriptionists},
  author={Zhou, Li and Blackley, Suzanne V and Kowalski, Leigh and Doan, Raymond and Acker, Warren W and Landman, Adam B and Kontrient, Evgeni and Mack, David and Meteer, Marie and Bates, David W and others},
  journal={JAMA network open},
  volume={1},
  number={3},
  pages={e180530--e180530},
  year={2018},
  publisher={American Medical Association}
}

@article{entwistle2010supporting,
  title={Supporting patient autonomy: the importance of clinician-patient relationships},
  author={Entwistle, Vikki A and Carter, Stacy M and Cribb, Alan and McCaffery, Kirsten},
  journal={Journal of general internal medicine},
  volume={25},
  number={7},
  pages={741--745},
  year={2010},
  publisher={Springer}
}

@article{xafis2019ethics,
  title={An ethics framework for big data in health and research},
  author={Xafis, Vicki and Schaefer, G Owen and Labude, Markus K and Brassington, Iain and Ballantyne, Angela and Lim, Hannah Yeefen and Lipworth, Wendy and Lysaght, Tamra and Stewart, Cameron and Sun, Shirley and others},
  journal={Asian Bioethics Review},
  volume={11},
  number={3},
  pages={227--254},
  year={2019},
  publisher={Springer}
}

@article{rockwern2021health,
  title={Health information privacy, protection, and use in the expanding digital health ecosystem: a position paper of the American College of Physicians},
  author={Rockwern, Brooke and Johnson, Dejaih and Snyder Sulmasy, Lois and Medical Informatics Committee and Ethics, Professionalism and Human Rights Committee of the American College of Physicians},
  journal={Annals of internal medicine},
  volume={174},
  number={7},
  pages={994--998},
  year={2021},
  publisher={American College of Physicians}
}

@article{ng2025evaluating,
  title={Evaluating the performance of artificial intelligence-based speech recognition for clinical documentation: a systematic review},
  author={Ng, Joel Jia Wei and Wang, Eugene and Zhou, Xinyan and Zhou, Kevin Xiang and Goh, Charlene Xing Le and Sim, Gabriel Zheng Ning and Tan, Hiang Khoon and Goh, Serene Si Ning and Ng, Qin Xiang},
  journal={BMC Medical Informatics and Decision Making},
  volume={25},
  number={1},
  pages={236},
  year={2025},
  publisher={Springer}
}

@article{goss2016incidence,
  title={Incidence of speech recognition errors in the emergency department},
  author={Goss, Foster R and Zhou, Li and Weiner, Scott G},
  journal={International journal of medical informatics},
  volume={93},
  pages={70--73},
  year={2016},
  publisher={Elsevier}
}

@inproceedings{tae2019data,
  title={Data cleaning for accurate, fair, and robust models: A big data-AI integration approach},
  author={Tae, Ki Hyun and Roh, Yuji and Oh, Young Hun and Kim, Hyunsu and Whang, Steven Euijong},
  booktitle={Proceedings of the 3rd international workshop on data management for end-to-end machine learning},
  pages={1--4},
  year={2019}
}

@article{daneshjou2021lack,
  title={Lack of transparency and potential bias in artificial intelligence data sets and algorithms: a scoping review},
  author={Daneshjou, Roxana and Smith, Mary P and Sun, Mary D and Rotemberg, Veronica and Zou, James},
  journal={JAMA dermatology},
  volume={157},
  number={11},
  pages={1362--1369},
  year={2021},
  publisher={American Medical Association}
}



\end{document}